\documentclass[onecolumn,conference]{IEEEtran}
\usepackage[final]{graphicx}
\usepackage{cite}
\usepackage{amssymb}
\usepackage{amsmath}
\usepackage{amsthm}
\usepackage{amsfonts}
\usepackage{bm}
\usepackage{filecontents}

\title{Robust Design of Power Minimizing Symbol-Level Precoder under Channel Uncertainty}

\author{Alireza~Haqiqatnejad,~Farbod~Kayhan,~and Bj\"{o}rn~Ottersten\\
	Interdisciplinary Centre for Security, Reliability and Trust (SnT),~University of Luxembourg \\
	email: \{alireza.haqiqatnejad,farbod.kayhan,bjorn.ottersten\}@uni.lu}

\newcommand{\Deee} {\mathrm{\boldsymbol{\delta}}}
\newcommand{\ups} {\mathrm{\boldsymbol{\upsilon}}}

\newcommand{\Del}{\mathrm{\mathbf{\Delta}}}
\newcommand{\Sigmaaa}{\mathrm{\mathbf{\Sigma}}}
\newcommand{\Siii}{\mathrm{\mathbf{\Psi}}}

\newcommand{\omg} {\mathrm{\boldsymbol{\omega}}}
\newcommand{\HHH} {\mathrm{\mathbf{H}}}
\newcommand{\bbb}{\mathrm{\mathbf{b}}}
\newcommand{\uuu}{\mathrm{\mathbf{u}}}

\newcommand{\WWW}{\mathrm{\mathbf{W}}}
\newcommand{\ZZZ}{\mathrm{\mathbf{Z}}}
\newcommand{\I}{\mathrm{\mathbf{I}}}
\newcommand{\J}{\mathrm{\mathbf{J}}}
\newcommand{\A}{\mathrm{\mathbf{A}}}

\newcommand{\T}{\mathrm{T}}
\newcommand{\F}{\mathrm{F}}
\newcommand{\HH}{\mathrm{H}}

\newcommand{\YYY}{\mathrm{\mathbf{Y}}}
\newcommand{\XXX}{\mathrm{\mathbf{X}}}
\newcommand{\OOO}{\mathrm{\mathbf{0}}}
\newcommand{\aaa}{\mathrm{\mathbf{a}}}
\newcommand{\h}{\mathrm{\mathbf{h}}}

\newcommand{\s}{\mathrm{\mathbf{s}}}

\newcommand{\ccc}{\mathrm{\mathbf{c}}}
\newcommand{\x}{\mathrm{\mathbf{x}}}

\newcommand{\EXP}{\mathrm{E}}
\newcommand{\PR}{\mathrm{Pr}}

\newcommand{\VEC}{\mathrm{vec}}

\newcommand{\INF}{\mathrm{inf}}
\newcommand{\TT}{\mathrm{T}}
\newcommand{\C}{\mathbb{C}}

\begin{document}
\maketitle

\begin{abstract}
In this paper, we investigate the downlink transmission of a multiuser multiple-input single-output (MISO) channel under a symbol-level precoding (SLP) scheme, having imperfect channel knowledge at the transmitter. In defining the SLP design problem, a general category of constructive interference regions (CIR) called distance preserving CIR (DPCIR) is adopted. In particular, we are interested in a robust SLP design minimizing the total transmit power subject to individual quality-of-service (QoS) requirements. We consider two common models for the channel uncertainty region, namely, spherical (norm-bounded) and stochastic. For the spherical uncertainty model, a worst-case robust precoder is proposed, while for the stochastically known uncertainties, we derive a convex optimization problem with probabilistic constraints. We simulate the performance of the proposed robust approaches, and compare them with the existing methods. Through the simulation results, we also show that there is an essential trade-off between the two robust approaches.
\end{abstract}


\section{Introduction}
Multiuser transmit beamforming is an effective way of improving the reliability and throughput of individual users in a multiuser wireless system. This advantage is brought by employing multiple antennas at the transmitter/receiver side, which enables the ability to manipulate the multiuser (channel-induced) interference through exploiting the spatial domain. In a multiuser downlink scenario, precoding schemes can be applied to mitigate the multiuser interference (MUI), and thus to enhance the performance, by spatially pre-processing the users' data stream; while at the same time, attempting to guarantee certain system-centric or user-specific requirements.

Multiuser precoding techniques typically exploit the channel knowledge in order to suppress the MUI \cite{tb_opt}. On the other hand, the notion of constructive interference (CI) has been introduced as a promising alternative where the underlying idea is to turn the MUI, which is often treated as an unwanted distortion, into a useful source of signal power \cite{slp_chr}. Following the CI-based design concept, the precoder's output is obtained on a symbol-by-symbol basis \cite{slp_chr, slp_con}, which is referred to as symbol-level precoding (SLP). Generally speaking, objective-oriented design of a multiuser precoder involves a constrained optimization problem aimed at finding the optimal transmit signal subject to provisioning a measure of the users' quality-of-service (QoS), e.g., the required signal-to-interference-plus-noise ratio (SINR) \cite{tb_convex}. Among a variety of multiuser precoding design criteria, a commonly addressed problem is QoS-constrained power minimization \cite{tb_vis,tb_sinr}, which is the primary focus of this paper.

The performance improvement promised by the multiuser precoding may not be realized if accurate channel state information (CSI) is not available at the transmitter. It is mainly because precoding schemes are quite sensitive to channel uncertainties \cite{unc_sens}. One may expect even a more adverse effect of imperfect channel knowledge when a symbol-level precoder is employed; this is due to the fact that the efficiency of the SLP design extremely depends on the satisfaction of CI constraints to properly accommodate the (noise-free) received signals in constructive interference regions (CIR). 
In reality, assuming perfect CSI, either statistical or instantaneous, is not practical due to various inevitable channel disturbances such as channel estimation errors, quantization errors, and latency-related errors \cite{unc_tract}.


Several robust approaches have been proposed in the literature on conventional multiuser precoding, mostly assuming a perturbation-based channel uncertainty. The uncertainty regions are usually considered to be spherical (see, e.g., \cite{unc_wc, unc_conic}), or stochastic (see, e.g., \cite{unc_sta, unc_imp}). In this context, the robustness generally means designing the precoder such that certain constraints are satisfied for all possible errors within the uncertainty region. 
Based on the spherical uncertainty model, the disturbance is supposed to be within a known norm-bounded uncertainty set, without any assumption on its distribution. This model, which ultimately leads to worst-case analysis, is known to appropriately capture the bounded disturbances resulted from quantization error \cite{tb_qos}.
Stochastic robustness, on the other hand, assumes known statistical properties for the channel uncertainty.
In scenarios with channel estimation at the receiver, such assumption may adequately characterize the perturbing component since the error in the estimation process can often be approximated as a Gaussian random variable \cite{tb_qos}. This model also enables handling the outage probability by replacing the worst-case constraints with probabilistic constraints \cite{outage_spec, outage_cri}.

In the literature on the SLP design, a worst-case robust analysis is presented in \cite{slp_chr} to design the symbol-level precoder with norm-bounded CSI errors, addressing the power minimization and SINR balancing problems. The symbol-level SINR balancing optimization approach with outage probability constraints is also reported in \cite{unc_chr} to achieve robustness against stochastic channel uncertainties. Both the aforementioned methods are restricted to PSK constellations in designing the robust precoder.

It is important to notice that as far as the power minimization problem is of concern, the spherical uncertainty model might not yield an efficient solution. This model considers the worst-case errors which inherently leads to increasing the transmit power, though enhancing the users' symbol error probability. In order to have a complete analysis of power minimizing precoders with imperfect channel knowledge, the study of stochastic models may be beneficial, which has not been addressed in the literature of SLP design. In this paper, we study the SLP power minimization problem with SINR constraints based on a general family of CIRs, namely, distance preserving CIRs (DPCIR) \cite{slp_gen}, in the presence of channel uncertainty. We consider both uncertainty models, i.e., norm-bounded spherical and stochastic, where the latter model is expected to better fit the nature of the power minimization problem. Under norm-bounded CSI uncertainties, we obtain a robust precoder taking the worst-case error into consideration. In the case where the statistical properties of the uncertainty is available, we design a stochastically robust precoder by defining a probabilistic (convex) optimization problem. We show that our proposed approach outperforms the existing results in the literature in terms of power efficiency of the precoding scheme.

The rest of this paper is organized as follows. In Section \ref{sec:sys}, we describe the system and uncertainty model. This is followed by Section \ref{sec:def}, where we define the DPCIR-based SLP power optimization problem . We then propose two robust formulations for the norm-bounded and stochastic uncertainty models in Section \ref{sec:slp}. Simulation results are provided in Section \ref{sec:sim}. Finally, we conclude the paper in Section \ref{sec:con}.

\noindent{\bf{Notation:}} We use uppercase and lowercase bold-faced letters to denote matrices and vectors, respectively. For complex scalars, $(\cdot)^*$ denotes the conjugate operator. For matrices and vectors, $[\,\cdot\,]^\HH$ and $[\,\cdot\,]^\T$ denote conjugate transpose and transpose operator, respectively. For a square matrix $\A$, $|\A|$ denotes the determinant of $\A$. For vectors, $\succeq$ denotes componentwise inequality. The operator $\VEC(\cdot)$ denotes vectorization, and $\mathrm{blkdiag}(\cdot)$ represents a square block matrix having main-diagonal block matrices and zero off-diagonal blocks. $\mathbf{I}$ stands for an identity matrix of appropriate dimension. The expectation operator is denoted by $\EXP\{\cdot\}$, and $\otimes$ denotes the Kronecker product.

\section{System and Uncertainty Model}\label{sec:sys}

We consider a downlink multiuser MISO (unicast) scenario in which a common transmitter, e.g., a base station (BS), sends independent data streams to $K$ single-antenna users. The BS is equipped with $N$ transmit antennas, and a frequency-flat fading channel is assumed between the BS's transmit antennas and any user $k$. The channel vectors are denoted by $\h_k\in\C^{1\times N}, k=1,...,K$, containing the complex channel coefficients. Independent data symbols $\{s_k\}_{k=1}^K$ are to be conveyed to $K$ users every symbol time, where $s_k$ denotes the intended symbol for the $k$-th user. To simplify the notation, the symbol's time index is dropped throughout the paper. The users' symbols $\{s_k\}_{k=1}^K$ are drawn from finite equiprobable two-dimensional constellation sets. Without loss of generality, we assume an identical $M$-ary constellation set with unit average power for all $K$ users. Collecting the users' symbols in a vector $\s=[s_1,\ldots,s_K]^\T\in\C^{K\times1}$, the symbol vector $\s$ is mapped onto $N$ transmit antennas through a symbol-level precoder \cite{slp_chr, slp_con}. This yields the output vector $\uuu=[u_1,\ldots,u_N]^\T\in\C^{N\times1}$ to be transmitted by the BS. The received signal by the $k$-th user is then equal to
\begin{equation}\label{eq:sys}
r_k = \h_k\uuu+z_k, \; k=1,...,K,
\end{equation}
where $z_k$ represents the additive complex Gaussian noise at the receiver of user $k$ with distribution $z_k\sim\mathcal{CN}(0,\sigma_k^2)$. We assume uncorrelated noise components across the receivers, i.e., $\EXP\{z_k z_j^*\} = 0, \forall k,j=1,...,K, k\neq j$. Having $r_k$, the $k$-th user may optimally detect its desired symbol $s_k$ based on the single-user maximum-likelihood (ML) decision rule.

We further consider a more realistic scenario in which the available channel knowledge at the BS is not accurate. A perturbation-based uncertainty is assumed according to which the $k$-th user's channel is equal to
\begin{equation}\label{eq:H}
\h_k = \hat{\h}_k + \Deee_k, \; k=1,...,K,
\end{equation}
where $\hat{\h}_k$ is the known erroneous channel associated with user $k$, and the perturbing component $\Deee_k\in\mathbb{C}^{1\times N}$ is characterized based on the adopted model for the uncertainty region. 
In the case of spherical uncertainty region, $\Deee_k$ is assumed to be a norm-bounded error vector, i.e.,
\begin{equation}\label{eq:delc}
\|\Deee_k\|_2 \leq \frac{1}{2}\delta_k, \; k=1,...,K,
\end{equation}
where $\delta_k$ specifies the radius of the uncertainty region related to the $k$-th user. On the other hand, considering a stochastic uncertainty, $\Deee_k$ represents a zero-mean Gaussian CSI error distributed as $\Deee_k\sim\mathcal{CN}(\OOO,2\,\xi_k^2\,\I)$. In both models, the random channel vectors $\{\h_k\}_{k=1}^K$ and the disturbances $\{\Deee_k\}_{k=1}^K$ are assumed to be mutually uncorrelated.

Hereafter, instead of complex-valued notations, we use the equivalent real-valued ones $\tilde{\uuu}=[\Re\{\uuu^\T\},\Im\{\uuu^\T\}]^\T\in\mathbb{R}^{2N\times1}$, $\HHH_k=\TT(\h_k)\in\mathbb{R}^{2\times2N}$, $\hat{\HHH}_k=\TT(\hat{\h}_k)\in\mathbb{R}^{2\times2N}$, and $\Del_k=\TT(\Deee_k)\in\mathbb{R}^{2\times2N},k=1,...,K$, where for any given complex-valued vector $\x$, the operator $\TT(\x)$ is defined as
\begin{equation}
\nonumber
\TT(\x)=
\begin{bmatrix}
\Re\{\x\} & -\Im\{\x\}\\
\Im\{\x\} & \Re\{\x\}
\end{bmatrix}.
\end{equation}
It is immediately apparent that $\|\Del_k\|_\F = 2\|\Deee_k\|_2$ and further
\begin{equation}\label{eq:Hr}
\HHH_k = \hat{\HHH}_k + \Del_k, \; k=1,...,K.
\end{equation}

In what follows, we simplify the norm notations such that $\|\cdot\|$ denotes either the Frobenius norm of a matrix or the Euclidean norm of a vector.

\section{Problem Definition}\label{sec:def}

We consider a design criterion based on which the symbol-level precoder is aimed at minimizing the total transmit power while guaranteeing certain SINR thresholds for the users. 
As mentioned before, the SLP design depends on the defined CIRs for any given constellation. In this paper, we adopt the DPCIRs introduced in \cite{slp_gen}, where the regions are defined such that the distances between the noise-free received signals are at least as large as the original distances of the constellation. 
Assuming DPCIRs and perfect CSI, the SLP design boils down to solving an SINR-constrained power minimization problem which has been expressed in \cite{slp_gen} as
\begin{equation}\label{eq:pm}
\begin{aligned}
\underset{\tilde{\uuu}}{\mathrm{min}} & \quad \tilde{\uuu}^\T\tilde{\uuu}\\
\mathrm{s.t.} & \quad \A_k \HHH_k \tilde{\uuu} \succeq \sigma_k\sqrt{\gamma_k}(\bbb_k+\ccc_k), \; k=1,...,K,
\end{aligned}
\end{equation}
where $\gamma_k$ is the given SINR threshold for user $k$, and $\A_k\in\mathbb{R}^{2\times2}$, $\bbb_k\in\mathbb{R}^2$ and $\ccc_k\in\mathbb{R}^2$ describe the hyperplane representation of the DPCIR associated with user $k$. In the rest, we define $\Siii_k\triangleq\sigma_k\sqrt{\gamma_k}(\bbb_k+\ccc_k)$. It should be noted that the optimization constraints in \eqref{eq:pm} accommodate each noise-free received signal $\HHH_k \tilde{\uuu}$ in its corresponding DPCIR, enhancing the detection accuracy and enlarging the feasibility region of the precoding design problem.

In the presence of channel uncertainty, the non-robust precoder solves the following optimization problem based on imperfect knowledge of the channel at the BS:
\begin{equation}\label{eq:pmnr}
\begin{aligned}
\underset{\tilde{\uuu}}{\mathrm{min}} & \quad \tilde{\uuu}^\T\tilde{\uuu}\\
\mathrm{s.t.} & \quad \A_k \hat{\HHH}_k \tilde{\uuu} \succeq \Siii_k, \; k=1,...,K.
\end{aligned}
\end{equation}
Nevertheless, for any user $k$, optimizing the transmit vector through \eqref{eq:pmnr} may cause imprecision in the location of the noise-free received signal due to the inaccurate channel $\hat{\HHH}_k$. More precisely, $\HHH_k \tilde{\uuu}$ may not be received in the intended DPCIR.
It is assumed that, in addition to the erroneous channel $\hat{\HHH}_k$, the BS is aware of either the error sphere radius $\delta_k$ or the statistics of $\Del_k$, for all the users $k=1,...,K$, depending on the adopted uncertainty model. In order to achieve robustness to CSI errors, we need to take their specifications into account when designing the symbol-level precoder. Accordingly, by exploiting our knowledge of $\{\Del_k\}_{k=1}^K$, in the next section our goal is to design the power minimizing symbol-level precoder being robust to partially known channel uncertainties.

\section{Robust Power Minimizing SLP Design}\label{sec:slp}

Having the perturbation-based uncertainty model in \eqref{eq:Hr}, the CI constraint for the $k$-th user can be written as
\begin{equation*}
\A_k (\hat{\HHH}_k + \Del_k) \tilde{\uuu} \succeq \Siii_k,
\end{equation*}
or equivalently
\begin{equation}\label{eq:ci}
\A_k \Del_k \tilde{\uuu} \succeq \Siii_k - \A_k \hat{\HHH}_k \tilde{\uuu}.
\end{equation}
In the sequel, we separately consider each uncertainty region and obtain the design formulation of the corresponding robust symbol-level precoder.

\subsection{Spherical Uncertainty Model}

The robust design of the precoder in this case aims at optimizing the transmit vector $\tilde{\uuu}$ while satisfying the constraints for any possible $\Del_k$ belonging to the region
\begin{equation}\label{eq:del}
\|\Del_k\| \leq \delta_k, \; k=1,...,K.
\end{equation}
This norm-bounded region can be interpreted as having all the errors inside a $2N$-dimensional sphere. Consequently, using \eqref{eq:ci} the power minimization problem is reformulated as
\begin{equation}\label{eq:pmall}
\begin{aligned}
\underset{\tilde{\uuu}}{\mathrm{min}} & \quad \tilde{\uuu}^\T\tilde{\uuu}\\
\mathrm{s.t.} & \quad \A_k \Del_k \tilde{\uuu} \succeq \Siii_k - \A_k \hat{\HHH}_k \tilde{\uuu}, \; k=1,...,K,\\
& \quad \forall \|\Del_k\| \leq \delta_k, \; k=1,...,K.
\end{aligned}
\end{equation}
In order to deal with norm-bounded CSI errors, a common approach is to design the precoder based on the worst-case uncertainty, which can be regarded as a conservative worst-case robustness \cite{unc_wc}. Accordingly, denoting $\A_k = [\aaa_{k,1},\aaa_{k,2}]^\T$ and $\Siii_k = [\psi_{k,1},\psi_{k,2}]^\T$, the optimization problem \eqref{eq:pmall} can be expressed as
\begin{equation}\label{eq:pmwc}
\begin{aligned}
\underset{\tilde{\uuu}}{\mathrm{min}} & \enspace \tilde{\uuu}^\T\tilde{\uuu}\\
\mathrm{s.t.} & \enspace \begin{bmatrix} \underset{\|\Del_k\| \leq \delta_k}{\INF} \! \{\aaa_{k,1}^\T \Del_k \tilde{\uuu}\} \\ \underset{\|\Del_k\| \leq \delta_k}{\INF} \! \{\aaa_{k,2}^\T \Del_k \tilde{\uuu}\} \end{bmatrix} \succeq \Siii_k - \A_k \hat{\HHH}_k \tilde{\uuu}, k=1,...,K.\\
\end{aligned}
\end{equation}
First, let focus on the first row of constraints in \eqref{eq:pmwc}. Using the property that for any given matrices $\XXX$, $\YYY$ and $\ZZZ$, we have $\VEC(\XXX\YYY\ZZZ)=(\ZZZ^\T \otimes \XXX) \VEC(\YYY)$ \cite{cookbook}, and that $\A_k \Del_k \tilde{\uuu}=\VEC(\A_k \Del_k \tilde{\uuu})$, the worst-case CI constraint for the $k$-th user is equivalent to
\begin{equation}\label{eq:wc}
\begin{aligned}
\underset{\|\Del_k\| \leq \delta_k}{\INF} \left\{(\tilde{\uuu}^\T \otimes \aaa_{k,1}^\T) \VEC(\Del_k) \right\} \geq \psi_{k,1} - \aaa_{k,1}^\T \hat{\HHH}_k \tilde{\uuu}.
\end{aligned}
\end{equation}
It can be shown that
\begin{equation}\label{eq:sup}
\begin{aligned}
\underset{\|\Del_k\| \leq \delta_k}{\INF} \left\{(\tilde{\uuu}^\T \otimes \aaa_{k,1}^\T) \VEC(\Del_k) \right\} = -\|\tilde{\uuu}^\T \otimes \aaa_{k,1}^\T\|\,\|\VEC(\Del_k)\| = -\delta_k \, \|\tilde{\uuu}^\T \otimes \aaa_{k,1}^\T\|,
\end{aligned}
\end{equation}
where the last equality is obtained having $\|\Del_k\| \! = \! \|\VEC(\Del_k)\|$. In fact, \eqref{eq:sup} accounts the worst possible case of the error $\Del_k$ by considering the maximal value of the inner product. Furthermore,
\begin{equation}\label{eq:norm}
\|\tilde{\uuu}^\T \otimes \aaa_{k,1}^\T\| = \|\tilde{\uuu}\| \, \|\aaa_{k,1}\|,
\end{equation}
which holds provided that both $\tilde{\uuu}$ and $\aaa_{k,1}$ are vectors.
Similar manipulation can be done for the second row of constraints in \eqref{eq:pmwc}.
Putting \eqref{eq:sup} and \eqref{eq:norm} together, problem \eqref{eq:pmwc} can be recast as
\begin{equation}\label{eq:pmwc2}
\begin{aligned}
\underset{\tilde{\uuu}}{\mathrm{min}} & \enspace \tilde{\uuu}^\T\tilde{\uuu}\\
\mathrm{s.t.} & \enspace \delta_k \, \|\tilde{\uuu}\| \, \left[\,\|\aaa_{k,1}\|,\|\aaa_{k,2}\|\,\right]^\T \! \preceq \! \A_k \hat{\HHH}_k \tilde{\uuu} \! - \! \Siii_k, k=1,...,K.\\
\end{aligned}
\end{equation}
This formulation ensures that the CI constraint for the $k$-th user will be met in the presence of any random, but norm-bounded CSI error. The robust formulation \eqref{eq:pmwc2} is a convex optimization problem and can efficiently be solved via off-the-shelf algorithms \cite{convex_boyd}.
A similar approach has also been studied in \cite{slp_chr} where the the CIRs coincide with the DPCIRs for $M$-PSK constellations, but the characterization of the CIRs are not identical. Therefore, presentation of the convex optimization problems are slightly different. Nevertheless, it should be noted that the final optimization problems are based on the same idea and are essentially equivalent. 

\subsection{Stochastic Uncertainty Model}

In case the BS knows the statistics of the channel perturbing components $\{\Del_k\}_{k=1}^K$, a reasonable approach is to design the precoder based on probabilistic constraints \cite{unc_sto}. In the context of SLP, this can be interpreted as considering probabilistic CI constraints \cite{unc_chr}. Accordingly, by modifying the deterministic constraints of the non-robust problem \eqref{eq:pmnr}, we define the stochastically robust power minimization as

\begin{equation}\label{eq:pmr}
\begin{aligned}
\underset{\tilde{\uuu}}{\mathrm{min}} & \quad \tilde{\uuu}^\T\tilde{\uuu} \\
\mathrm{s.t.} & \quad 1 \!-\! \PR\left\{\A_k \Del_k \tilde{\uuu} \succeq \Siii_k - \A_k \hat{\HHH}_k \tilde{\uuu}\right\} \leq \epsilon, k=1,...,K,
\end{aligned}
\end{equation}
where $\epsilon$ denotes the threshold probability that allows the noise-free received signal to locate outside the intended DPCIR. In other words, the precoder is so designed to keep the probabilities of violating the CI constraints below a certain value. The probabilistic constraint in \eqref{eq:pmr}, for any $k=1,...,K$, can be written as
\begin{equation}\label{eq:prob}
\PR\left\{(\tilde{\uuu}^\T \otimes \A_k) \VEC(\Del_k) \succeq \Siii_k - \A_k \hat{\HHH}_k \tilde{\uuu}\right\} \geq 1-\epsilon.
\end{equation}
The vector of Gaussian CSI errors, $\VEC(\Del_k)$, is characterized by its mean and covariance matrix given by
\begin{equation*}
\EXP\{\VEC(\Del_k)\} = \; \OOO,
\end{equation*}
and
\begin{equation}\label{Edel}
\EXP\{\VEC(\Del_k)\VEC(\Del_k)^\T\} = \xi_k^2\begin{bmatrix} \I & \J \\
\J^\T & \I
\end{bmatrix},
\end{equation}
respectively, where $$\J=\mathrm{blkdiag}(\J_1,...,\J_N),\J_n =  \begin{bmatrix} 0 & 1 \\ -1 & 0
\end{bmatrix},\forall n\in\{1,...,N\}.$$
Next, let $\ups_k \triangleq (\tilde{\uuu}^\T \otimes \A_k) \; \VEC(\Del_k) = [\upsilon_{k,1},\upsilon_{k,2}]^\T$ and $\omg_k \triangleq \Siii_k - \A_k \hat{\HHH}_k \tilde{\uuu} = [\omega_{k,1},\omega_{k,2}]^\T$. By definition, $\omg_k$ is a deterministic function of $\tilde{\uuu}$, and $\ups$ is a bivariate Gaussian random variable (r.v.) which is characterized by
\begin{equation*}
\begin{aligned}
\EXP\{\ups_k\} & = \EXP\left\{(\tilde{\uuu}^\T \otimes \A_k) \; \VEC(\Del_k)\right\} \\
& = (\tilde{\uuu}^\T \otimes \A_k) \; \EXP\left\{\VEC(\Del_k)\right\} = \OOO,
\end{aligned}
\end{equation*}
and
\begin{equation}\label{sigk}
\begin{aligned}
\Sigmaaa_k &= \EXP\{\ups_k \ups_k^\T\}\\
&= \EXP\left\{\left((\tilde{\uuu}^\T \! \otimes \! \A_k) \VEC(\Del_k)\right)\!\left((\tilde{\uuu}^\T \! \otimes \! \A_k) \VEC(\Del_k)\right)^\T\right\}\\
& \overset{\mathrm{(a)}}{=} (\tilde{\uuu}^\T \otimes \A_k)\EXP\left\{\VEC(\Del_k)\VEC(\Del_k)^\T\right\}(\tilde{\uuu} \otimes \A_k^\T) \;\\
& \overset{\mathrm{(b)}}{=} \xi_k^2 \, (\tilde{\uuu}^\T \otimes \A_k)(\tilde{\uuu} \otimes \A_k^\T) \;\\
& \overset{\mathrm{(c)}}{=} \xi_k^2 \, (\tilde{\uuu}^\T \tilde{\uuu} \otimes \A_k \A_k^\T)\\
& = \xi_k^2 \, \|\tilde{\uuu}\|^2 \, \A_k \A_k^\T,
\end{aligned}
\end{equation}
where the equalities (a) and (c) are respectively derived by applying the properties $(\XXX \otimes \YYY)^\T=(\XXX^\T \otimes \YYY^\T)$ and $(\XXX \otimes \YYY)(\WWW \otimes \ZZZ)=(\XXX \WWW \otimes \YYY \ZZZ)$, for any given matrices $\XXX,\YYY,\WWW,\ZZZ$ \cite{cookbook}. Furthermore, the equality (b) in \eqref{sigk} can be easily verified using \eqref{Edel}, however, the intermediate steps are omitted for brevity.
Having the statistics of $\ups_k$, the left-hand side probability in \eqref{eq:prob} is computed by integrating the joint Gaussian density function of $\upsilon_{k,1}$ and $\upsilon_{k,2}$, i.e.,
\begin{equation}\label{eq:int}
\begin{aligned}
\PR \{\ups_k \succeq \omg_k\} &= \PR\left\{\upsilon_{k,1} \geq \omega_{k,1},\upsilon_{k,2} \geq \omega_{k,2}\right\}\\
&= \int\limits_{\omega_{k,2}}^{\infty}\int\limits_{\omega_{k,1}}^{\infty}\frac{1}{2\pi\sqrt{|\Sigmaaa_k|}} \exp\left\{-\frac{1}{2}\ups_k^\T\Sigmaaa_k^{-1}\ups_k\right\} \mathrm{d}\upsilon_{k,1} \mathrm{d}\upsilon_{k,2}.
\end{aligned}
\end{equation}
This integration, however, has no closed-form expression and it becomes even more difficult to handle when included as a constraint into the problem \eqref{eq:pmr}, since it is a function of the optimization variable $\tilde{\uuu}$. To tackle this challenge, we apply a decorrelation transform \cite{papoulis} to the Gaussian random vector $\ups_k$ in order to find a tractable expression for \eqref{eq:int}. Denoting $\bar{\omg}_k=(\A_k \A_k^\T)^{-1/2}\omg_k=[\bar{\omega}_{k,1},\bar{\omega}_{k,2}]^\T$, we obtain
\begin{equation}\label{eq:dec}
\begin{aligned}
\PR \left\{\ups_k \succeq \omg_k\right\} &= \PR \left\{\Sigmaaa_k^{1/2}\Sigmaaa_k^{-1/2}\ups_k \succeq \omg_k\right\}\\
&= \PR \left\{\Sigmaaa_k^{1/2}\bar{\ups}_k \succeq \omg_k\right\}\\
&= \PR \left\{\bar{\ups}_k \succeq \Sigmaaa_k^{-1/2}\omg_k\right\}\\
&= \PR \left\{\bar{\ups}_k \succeq \frac{\bar{\omg}_k}{\xi_k \, \|\tilde{\uuu}\|}\right\},
\end{aligned}
\end{equation}
where the decorrelating matrix $\Sigmaaa_k^{-1/2}$ is the inverse square root of $\Sigmaaa_k$, and $\bar{\ups}_k=\Sigmaaa_k^{-1/2}\ups_k=[\bar{\upsilon}_{k,1},\bar{\upsilon}_{k,2}]^\T$ is an uncorrelated bivariate Gaussian r.v. with zero mean and unit diagonal covariance, i.e.,
\begin{equation}\label{eq:white}
\begin{aligned}
\bar{\Sigmaaa}_k &= \EXP\left\{\bar{\ups}_k \bar{\ups}_k^\T\right\}\\
&= \EXP\left\{\Sigmaaa_k^{-1/2}\ups_k \ups_k^\T \Sigmaaa_k^{-1/2}\right\}\\
&= \Sigmaaa_k^{-1/2}\EXP\left\{\ups_k \ups_k^\T\right\}\Sigmaaa_k^{-1/2}\\
&= \Sigmaaa_k^{-1/2} \Sigmaaa_k \Sigmaaa_k^{-1/2} = \I.
\end{aligned}
\end{equation}
Notice that since $\Sigmaaa_k$ is positive semidefinite, it has a unique square root. Furthermore, according to \cite{slp_tsp}, $\A_k \A_k^\T$ is always invertible. This implies the non-singularity of $(\A_k \A_k^\T)^{1/2}$, and hence the existence and uniqueness of $\Sigmaaa_k^{-1/2}$. It then follows from \eqref{eq:dec} and \eqref{eq:white} that
\begin{equation}\label{eq:erf}
\begin{aligned}
\PR \left\{\ups_k \succeq \omg_k\right\} &= \PR \left\{\bar{\upsilon}_{k,1} \geq \frac{\bar{\omega}_{k,1}}{\xi_k\|\tilde{\uuu}\|}\right\} \, \PR \left\{\bar{\upsilon}_{k,2} \geq \frac{\bar{\omega}_{k,2}}{\xi_k\|\tilde{\uuu}\|}\right\}\\
&= \left(\frac{1}{2} - \frac{1}{2} \mathrm{erf}\left(\frac{\bar{\omega}_{k,1}}{\sqrt{2}\,\xi_k\|\tilde{\uuu}\|}\right)\right) \left(\frac{1}{2} - \frac{1}{2} \mathrm{erf}\left(\frac{\bar{\omega}_{k,2}}{\sqrt{2}\,\xi_k\|\tilde{\uuu}\|}\right)\right),
\end{aligned}
\end{equation}
where $\mathrm{erf}(\cdot)$ is the Gauss error function. Due to the increasing monotonicity of the error function, the probability \eqref{eq:erf} is lower bounded by
\begin{equation}\label{eq:erflb}
\begin{aligned}
\PR \{\ups_k \succeq \omg_k\} &\geq \left(\frac{1}{2} - \frac{1}{2} \mathrm{erf}\left(\frac{\max\{\bar{\omega}_{k,1},\bar{\omega}_{k,2}\}}{\sqrt{2}\,\xi_k\|\tilde{\uuu}\|}\right)\right)^2.
\end{aligned}
\end{equation}
Considering \eqref{eq:erflb}, the probabilistic constraint \eqref{eq:prob} simplifies to
\begin{equation}\label{eq:prob2}
\begin{aligned}
\left(\frac{1}{2} - \frac{1}{2} \mathrm{erf}\left(\frac{\max\{\bar{\omega}_{k,1},\bar{\omega}_{k,2}\}}{\sqrt{2}\,\xi_k\|\tilde{\uuu}\|}\right)\right)^2 \geq 1-\epsilon,
\end{aligned}
\end{equation}
from which the subsequent steps to obtain the following linear inequality is straightforward
\begin{equation}\label{eq:lmi}
\sqrt{2}\rho(\epsilon)\xi_k\|\tilde{\uuu}\| \geq \max\{\bar{\omega}_{k,1},\bar{\omega}_{k,2}\},
\end{equation}
with $\rho(\epsilon) \triangleq \mathrm{erf}^{-1}\left(1 - 2 \sqrt{1 - \epsilon}\right)$.

\noindent Using \eqref{eq:lmi}, the robust power minimization can be formulated as a convex optimization problem expressed by
\begin{equation}\label{eq:pmrlmi}
\begin{aligned}
\underset{\tilde{\uuu}}{\mathrm{min}} & \quad \tilde{\uuu}^\T\tilde{\uuu}\\
\mathrm{s.t.} &\quad\!\max\left\{(\A_k \A_k^\T)^{-1/2}(\Siii_k - \A_k \hat{\HHH}_k \tilde{\uuu})\right\}\!\leq \!\sqrt{2}\rho(\epsilon)\xi_k\|\tilde{\uuu}\|,\\
& \quad k=1,...,K,
\end{aligned}
\end{equation}
which can be solved via several efficient methods known in the literature of convex optimization theory \cite{convex_boyd}. It is worth noting that the inequality \eqref{eq:lmi} is a stricter constraint than \eqref{eq:prob}, which is a consequence of using the probability lower bound \eqref{eq:erflb}. Therefore, the optimal solution of \eqref{eq:pmrlmi} is an upper bound on the optimum of \eqref{eq:pmr}, i.e., on the lowest possible transmit power for the stochastically robust precoder.

\section{Simulation Results}\label{sec:sim}

In this section, we provide some simulation results to evaluate the performance of different robust SLP approaches. The simulations have been done using MATLAB software and CVX convex optimization package (SDPT3 solver). In all the simulations, we consider a downlink multiuser MISO channel with $N=K=4$, in which the intended symbols of all the users are taken from an 8-PSK constellation. A Rayleigh block fading channel is assumed between any user $k$ and the BS's antennas where the channel coefficients are generated following an i.i.d. complex Gaussian distribution, i.e., $\h_k\sim\mathcal{CN}(\OOO,\I)$. It is further assumed that $\EXP\{\h_k^\HH\h_j\}=\OOO, \forall k,j=1,...,K,k\neq j$. We consider unit noise variance at the receiver of all the users, and also equal SINR thresholds, i.e., $\gamma_k=\gamma,k=1,...,K$. The threshold probability $\epsilon$ is set to $0.01$, unless otherwise stated. In the presence of stochastic uncertainties, zero-mean Gaussian CSI errors of equal variances $\xi_k^2=\xi^2$ are assumed for all the users. In order to have a rather fair comparison between the two uncertainty models, the norm-bounded CSI errors are so generated to be within balls of equal radii $\delta_k=\delta=\sqrt{2N}\xi,k=1,...,K$, where the errors are uniformly chosen from the uncertainty sets. For the spherical uncertainty model, we only present the results obtained from the equivalent worst-case robust symbol-level precoder in \cite{slp_chr}. All the plots are obtained by averaging the results over $200$ fading blocks each of $50$ symbol slots. In order for the results to be interpretable, the same set of channel realizations is considered for all SINR thresholds. In a more extensive study, one needs to differently generate the channel matrix for each SINR and solves the optimization problems for all possible combinations of the users' symbols. However, by doing so, extensive simulations through thousands of channel realizations with symbol slots of order $M^K$ are required to have reliable statistics.

\begin{figure}
	\centering
	\includegraphics[trim={0 0 0 .25in},clip,width=.5\columnwidth]{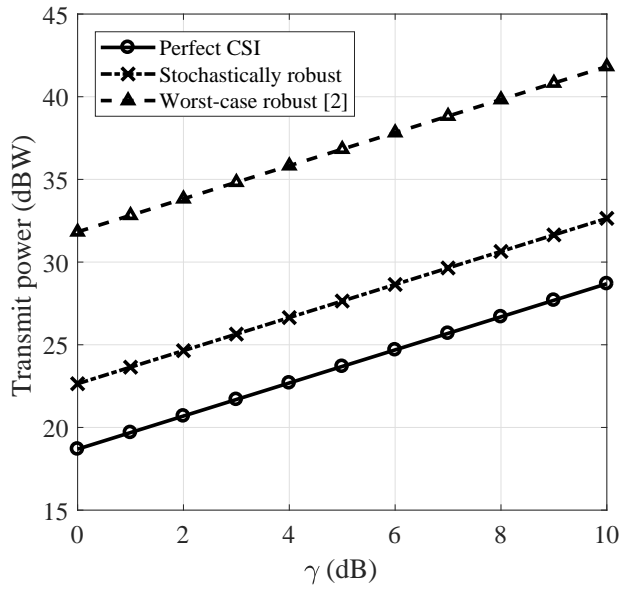}
	\caption{Optimized total transmit power versus SINR threshold.}
	\label{fig:1}
\end{figure}

\begin{figure}
	\centering
	\includegraphics[width=.5\columnwidth]{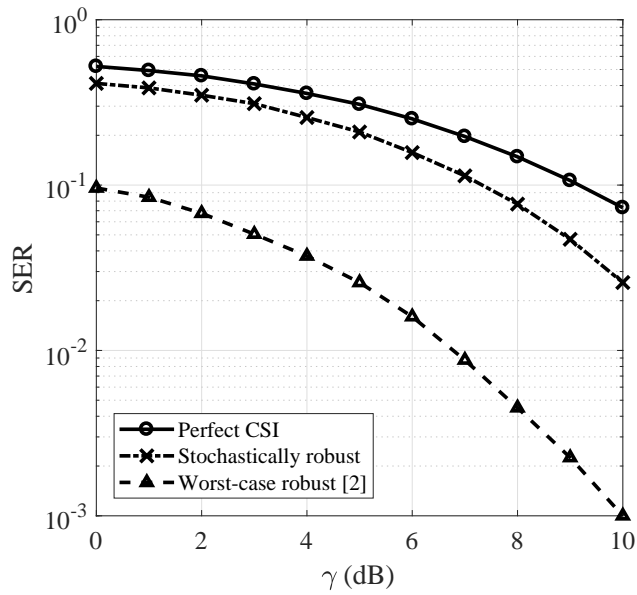}
	\caption{Average users' symbol error probability versus SINR threshold.}
	\label{fig:2}
\end{figure}

\begin{figure}
	\centering
	\includegraphics[width=.5\columnwidth]{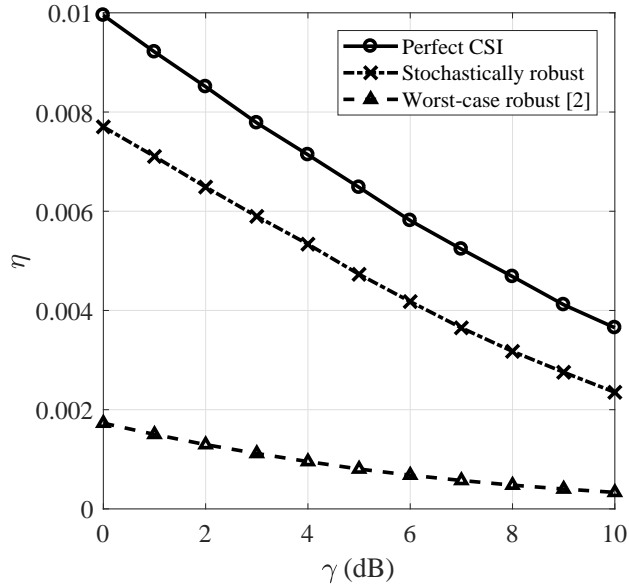}
	\caption{Power efficiency as a function of SINR threshold.}
	\label{fig:3}
\end{figure}

Figures \ref{fig:1} and \ref{fig:2} show the optimized transmit power and the average users' symbol error rate (SER), respectively, versus the SINR threshold. It can be observed from Fig. \ref{fig:1} that the proposed stochastically robust precoder reduces the transmit power by around $9$ dBW compared to the worst-case scheme. Moreover, robust precoding schemes lead to higher transmit powers when compared to the case with perfect channel knowledge which is an expected cost in order to achieve robustness. On the other hand, the worst-case robust scheme results in lower average SERs with an approximate gain of $7$ dB compared to the stochastically robust method, as can be seen in Fig. \ref{fig:2}, which is of course due to higher power consumption. This, however, means that the users are provided with higher SINRs than the required QoS levels, which may not be efficient in general, especially when the goal is to minimize the transmit power under a given SER requirement. Therefore, in order to make a more meaningful comparison between the overall performance of the two categories of robust precoding schemes, i.e., the worst-case and the stochastic, we define the power efficiency $\eta$ as the ratio between the average per-user throughput and the total transmit power, i.e.,
\begin{equation}\label{eq:pe} \nonumber
\eta = \frac{\frac{1}{K} \sum_{k=1}^K (1-\mathrm{SER}_k) \log_2(1+\|\HHH_k\tilde{\uuu}\|^2)}{\|\tilde{\uuu}\|^2}.
\end{equation}
In Fig. \ref{fig:3}, the power efficiency of different robust approaches is plotted versus the SINR threshold. The ratio $\eta$ can be interpreted as a trade-off factor between the achievable throughput (as a function of the SER performance) and the required transmit power. It can be seen that the stochastic-based power minimization scheme provides a more power-efficient robustness to known channel uncertainties.

\section{Conclusion and Future Research}\label{sec:con}

In this paper, we study the robust design of SLP under CSI uncertainty. First, we formulate the power minimization problem considering two different models for the uncertainty, namely, spherical and stochastic. For the minimum power SLP design with spherical (norm-bounded) CSI errors, which has been previously addressed for $M$-PSK constellations, we provide a generic formulation based on DPCIRs. Moreover, we formulate the problem also for the stochastic uncertainty model, where the noise-free received signals are allowed to not fall within the DPCIRs with a given probability. The main challenge in this case is that the probabilistic constraints in the optimization problem are not easy to handle. We use a rather efficient simplification that allows us to obtain convex constraints after decorrelating the error components. Our simulations show that there is an essential trade-off between the two robust approaches. While the worst-case method may always provide the requested target SER, the transmit power is usually increased considerably. On the other hand, in the stochastic approach, the increase in the transmit power with respect to the scenario with perfect CSI is quite smaller ($\approx 4$ dBW), however, it leads to higher SERs compared to the worst-case scheme. We further discuss the systems in which a target SER is needed to be satisfied, and introduce the power efficiency of the system as a comparison ratio. The power efficiency incorporates both the total transmit power and the average per-user throughput. The simulation results indicate that the power efficiency of the stochastic model is much higher with respect to that of the worst-case analysis.
An interesting problem to be investigated is to understand this trade-off for other types of modulation scheme. In addition, addressing the SLP design for SINR balancing and sum-rate maximization problems under given uncertainty models can be a topic of future work. 

\section*{Acknowledgment}
The authors are supported by the Luxembourg National Research Fund (FNR) under CORE Junior project: C16/IS/11332341 Enhanced Signal Space opTImization for satellite comMunication Systems (ESSTIMS).


\end{document}